\documentstyle[epsf,prd,aps]{revtex}
\begin{document}
\draft
\title
{Readings of the Lichnerowicz - York equation}
\author
{ Niall \'O Murchadha}
\address {Physics Department, University College, Cork, Ireland.}
\address {email: niall@ucc.ie}
\maketitle
\begin{abstract}
James York, in a major extension of Andr\'e Lichnerowicz's work, showed how to construct
solutions to the constraint equations of general relativity. The York method consists of
choosing a 3-metric on a given manifold; a divergence-free, tracefree (TT)
symmetric 2-tensor wrt this metric; and a single number, the trace of the extrinsic
curvature. One then obtains a quasi-linear elliptic equation for a scalar function,
the Lichnerowicz-York (L-Y) equation. The solution of this equation is used as a
conformal factor to transform the data into a set that satisfies the constraints. If the
manifold is compact and without boundary, one quantity that emerges is the volume
of the physical space. This article reinterprets the L-Y equation as an
eigenvalue equation so as to get a set of data with a preset physical volume. One 
chooses the conformal metric, the TT tensor, and the physical volume, while regarding the
trace of the extrinsic curvature as a free parameter. The resulting equation has
extremely nice uniqueness and existence properties. A even more radical approach would be
to fix the base (conformal) metric, the physical volume, and the trace.  One also selects
a TT tensor, but one is free to multiply it by a constant(unspecified). One then solves
the L-Y equation as an eigenvalue equation for this constant. A third choice would
be to fix the TT tensor and and multiply the base metric by a constant. Aech of these
three formulations has good uniqueness and existence properties.
\end{abstract}

\section{Introduction}

When general relativity is considered as a dynamical system, the Einstein equations,
just like the Maxwell equations, split into constraint and evolution
equations. In this article I focus entirely on the constraints. The initial data
consists of a pair $(g_{ij}, K^{ij})$, defined on a given 3-manifold, where $g_{ij}$ is a
Riemannian 3-metric and $K^{ij}$ is a symmetric tensor. $K^{ij}$ is the extrinsic
curvature of the 3-manifold when regarded as an embedded hypersurface of a
pseudo-Riemannian 4-manifold that satisfies the Einstein equations. \cite{adm}

The initial data cannot be specified freely: they must satisfy both a scalar and vector
equation, known respectively as the Hamiltonian and momentum constraints, i.e.,
\begin{equation}
R - K_{ij}K^{ij} + K^2 = 0, \label{ham}
\end{equation}
\begin{equation}
\nabla_i(K^{ij} - g^{ij}K) = 0, \label{mom}
\end{equation}
where $R$ is the scalar curvature and $K = g_{ij}K^{ij}$ is the trace of the extrinsic
curvature. This trace $K$ will play a key role in the rest of this article.

I (implicitly) define the extinsic curvature by writing the relation between the
extrinsic curvature and the time derivative of the metric as
\begin{equation}
\partial_t g_{ij} = 2NK_{ij} + \nabla_iN_j + \nabla_jN_i, \label{gdot}
\end{equation}
where $N$ is the lapse and $N^i$ is the shift.
This means that my sign convention for $K^{ij}$ agrees with Wald \cite{wald} and is the
opposite of Misner, Thorne, Wheeler \cite{mtw}. In particular 
\begin{equation}
{1 \over \sqrt{g}}\partial_t \sqrt{g} = NK + \nabla_iN^i. \label{K}
\end{equation}

The first significant step in understanding how to solve the constraints was
achieved by Andr\'e Lichnerowicz in 1944 \cite{lich}. He realised that if the extrinsic
curvature was tracefree, then the momentum constraint, Eq.(\ref{mom}), required that the
extrinsic curvature be both divergence-free and tracefree (described as transverse
tracefree, hence TT). TT tensors are conformally covariant. Given a TT
tensor wrt a given metric, on making the following transformations
\begin{equation}
\overline{g_{ij}} = \phi^4 g_{ij},\hskip 1cm \overline{K^{ij}_{TT}} =
\phi^{-10}K^{ij}_{TT},
\label{TT}
\end{equation}
then $\overline{K^{ij}_{TT}}$ is TT wrt the transformed metric $\overline{g_{ij}}$.
Thus it continues to satisfy the tracefree momentum constraint. Lichnerowicz had the idea
of choosing a base metric and a TT tensor on it, and making a conformal transformation
so that the transformed metric would satisfy the Hamiltonian constraint, i.e.,
\begin{equation}
\overline{R} - \overline{K^{ij}_{TT}}\overline{K^{TT}_{ij}} = 0. \label{Hbar}
\end{equation}
This reduces to the well-known Lichnerowicz equation for the conformal factor $\phi$
\begin{equation}
8\nabla^2\phi - R\phi + K^{ij}_{TT}K^{TT}_{ij}\phi^{-7} = 0.
\label{lich}
\end{equation}
Of course, since the solution $\phi$ is to be used as a conformal factor, one is only
interested in solutions which are everywhere positive. 

It turns out that for either the
asymptotically flat case or for the compact, without boundary, case, the Lichnerowicz
equation can only be solved for a restricted class of data.  The
choice of TT data is essentially irrelevant, but the choice of the base metric is
important. The `tracefree' Hamiltonian constraint Eq.(\ref{Hbar}) requires the solution
scalar curvature to be everywhere nonnegative. There exists a conformal invariant (the
Yamabe constant) \cite{ya}
\begin{equation}
Y = \inf {\int R \theta^2 + 8(\nabla \theta)^2 dv \over [\int \theta^6 dv]^{1 \over 3}},
\label{Y}
\end{equation}
where the infimum is taken over smooth functions in the compact case and over smooth
functions of compact support in the asymptotically flat case. The minimizing function
solves the Yamabe equation, and, when used as a conformal factor, transforms the
manifold into a compact one with constant scalar curvature. On such a manifold, the
minimizing function can be chosen equal to 1, and then we get
\begin{equation}
Y = {R_0V_0 \over V_0^{1 \over 3}} = R_0V_0^{2 \over 3}\label{Y'}.
\end{equation}
A given metric can be
conformally transformed into a metric with positive scalar curvature if and only if the
Yamabe constant is positive. As a result, the Lichnerowicz equation can be solved if and
only if the metric belongs to the positive Yamabe class.

 In 1970, James York \cite{y} extended the Lichnerowicz approach significantly when he
realised that one could add a constant trace term, $K$, to the extrinsic curvature
without disturbing the conformal covariance of the momentum constraint, i.e.,
\begin{equation}
\overline{g_{ij}} = \phi^4 g_{ij},\hskip 1cm \overline{K^{ij}} = \phi^{-10}K^{ij}_{TT}
+ {1
\over 3}K\overline{g^{ij}}.
\label{TTK}
\end{equation}
Note that the constant trace term is a conformal invariant! The equation for the
conformal factor becomes the Lichnerowicz - York equation
\begin{equation}
8\nabla^2\phi - R\phi + K^{ij}_{TT}K^{TT}_{ij}\phi^{-7} - {2 \over 3}K^2\phi^5 = 0.
\label{ly}
\end{equation}
The Hamiltonian constraint no longer places any restriction on the sign of the scalar
curvature when the trace of the extrinsic curvature is nonzero. Therefore the
Lichnerowicz - York equation can almost invariably be solved with any choice of base
metric, any choice of TT tensor, and any choice of $K$. It works both in the compact and
the asymptotically flat cases. The initial data that is constructed has constant trace of
the extrinsic curvature and is thus called a constant mean curvature (CMC) solution to
the constraints.

The standard method is quite straightforward: one chooses the base metric, a TT tensor,
and a value of $K$; one then solves for the conformal factor and constructs the physical
metric; given the physical metric, one can find the physical volume.
The aim of this article, however, is to show that the `standard' interpretation of the
Lichnerowicz - York equation, at least in the compact
case, is not the only reasonable one.  It turns out that one can trade off
knowledge of
$K$ for a specification of the physical volume ($V_p = \int \phi^6 dv$). This means
reading Eq.(\ref{ly}) as an eigenvalue equation for
$K$ by putting a normalization condition on the conformal factor so that the physical
volume has a prespecified value.

There are even more radical interpretations. If one is given a tensor which is TT wrt
some metric, then one can multiply it by any constant and it remains TT. Now solve the
Lichnerowicz - York equation, specifying the base metric, the TT tensor defined up to a
multiplicative constant, the value of $K$, and the physical volume.  The freedom to
choose the factor multiplying the TT tensor means that this is also a
wellposed system.

Finally, one could fix the TT tensor, but allow the base metric be defined up to a
multiplicative constant. Again, both $K$ and $V$ can be fixed and the constant regarded
as an eigenvalue. These three options will be discussed further.

\section{Trading off $K$ for $V_p$}

As part of his analysis of the constraints, James York \cite{y} suggested that
conformal superspace - the space of Riemannian 3-metrics, modulo diffeomorphisms, and
conformal transformations \cite{CS} - should be regarded as the natural configuration
space of GR. Recently, a number of us \cite{fabkom} have  found an action on conformal
superspace which generates GR in the CMC gauge.  A major difference here is that we need
to specify the physical volume as part of our initial data, while the value of $K$
emerges as part of the solution. An obvious question is whether this switch loses
us the extremely nice existence and uniqueness properties of the original form of the
Lichnerowicz - York equation. The answer, as revealed here, is that we lose essentially
nothing. 

As can be seen from Eq.(\ref{K}), and as previously pointed out by York \cite{y}, $K$
and the volume can be regarded as canonically conjugate variables; therefore
switching from $K$ to $V_p$ should be viewed as a Legandre transformation, and so should
not cause any  major disruption of the system. In \cite{fabkom} we write down an action,
so we are working in the Lagrangian rather that the Hamiltonian framework. Therefore it
seems more natural to use the volume (a metric quantity) rather than specifying $K$,
which is part of the conjugate momentum. Nevertheless, these sorts of arguments can only
be regarded as poor substitutes for showing that the equation itself is well behaved
after the switch from $K$ to $V_p$.

What we need to do is to fix the (conformal) metric and the TT tensor, and to
allow $K^2$ to span the entire range from 0 to $\infty$, solving the Lichnerowicz - York
equation for each value of $K^2$. We should then investigate what happens to the physical
volume as a function of
$K^2$. Note that the sign of $K$ does not matter in the Lichnerowicz - York equation:
only $K^2$ appears. The sign plays a role only in the evolution, when we determine
whether the universe either expands or contracts to the future. Let us start with a
perturbation calculation.

We begin with a solution to the
constraints, i.e., we are given a metric, a TT tensor, and a constant $K$, which satisfy
\begin{equation}
R - K_{ij}^{TT}K^{ij}_{TT}  + {2 \over 3}K^2 = 0. \label{ham1}
\end{equation}
We perturb this data by changing only $K^2$, by an amount $\delta K^2$, and
solve the perturbed Lichnerowicz - York equation (remember we are perturbing around
$\phi \equiv 1$)
\begin{equation}
8\nabla^2\delta\phi - R\delta\phi -7 K^{ij}_{TT}K^{TT}_{ij}\delta\phi - {10 \over
3}K^2\delta\phi = {2 \over 3}\delta K^2.
\label{ly1}
\end{equation}
This is an inhomogeneous linear elliptic equation for $\delta \phi$.
We can simplify it by eliminating $R$ using Eq.(\ref{ham1}) to get
\begin{equation}
8\nabla^2\delta\phi  -(8 K^{ij}_{TT}K^{TT}_{ij} + {8 \over
3}K^2) \delta\phi = {2 \over 3}\delta K^2.
\label{ly2}
\end{equation}
This is a particularly nice elliptic equation, because the coefficient of the linear term
is everywhere negative, which, in turn, guarantees that a unique solution to the
inhomogeneous equation exists. This should not come as a surprise, because we know that
the nonlinear equation always has a regular solution. What is interesting, however, is
that we can use the maximum principle to control the sign of $\delta \phi$. Let us
assume that $\delta K^2$ is positive. If $\delta \phi$ were positive in any region, we
would have a point where $\delta \phi$ has a positive maximum. However, at a
positive maximum, $\nabla^2\delta\phi$ is nonpositive, and the linear term is negative,
which is incompatible with a positive right hand side.  Therefore, we know that $\delta
\phi < 0$. This means that as $K^2$ increases $\phi$ decreases at each point, and the
physical volume monotonically decreases. The mapping between $V_p$ and $K^2$ is one to
one.

We have a fixed base metric,
a fixed TT tensor, and a $K^2$ which can become either unboundedly large or go to zero.
The next question to ask is what happens when $K^2$ becomes large. It can be shown
that $\phi$ shrinks monotonically to 0 everywhere, and the physical volume vanishes.
The easiest way to see this is to consider the alternative.  For each choice of $K^2$
we solve for $\phi$, which we know decreases. Let us assume that this $\phi$ does not
go to zero everywhere in the limit. In the region where $\phi$ is bounded away from zero,
the physical scalar curvature and the physical TT tensor remain bounded, while $K^2$
blows up. This is not compatible with solving the Hamiltonian constraint. Therefore
$\phi$ shrinks everywhere to zero, and the physical volume does likewise.

We can do better.
For ease of computation let us first make a conformal transformation so that the base
metric has constant scalar curvature. The Lichnerowicz - York equation then becomes
\begin{equation}
8 \nabla^2\phi - R_0\phi + K^{ij}_{TT}K^{TT}_{ij}\phi^{-7} - {2 \over 3}K^2\phi^5 = 0.
\label{ly0}
\end{equation}
At  the maximum of $\phi$ we have $\nabla^2\phi \le 0$. At this point we have
\begin{equation}
 K^{ij}_{TT}K^{TT}_{ij} \ge R_0\phi^8_{max} + {2 \over 3}K^2\phi^{12}_{max}. \label{in}
\end{equation}
The $K^{TT}K_{TT}$ term is evaluated at the point where $\phi$ reaches its maximum. We
can replace this in inequality (\ref{in}) by the maximum value over the whole manifold to
give
\begin{equation}
 K^{ij}_{TT}K^{TT}_{ij}(max) \ge R_0\phi^8_{max} + {2 \over 3}K^2\phi^{12}_{max}.
\label{in'}
\end{equation}
This means that as $K^2$ increases $\phi_{max}$ must decrease, and therefore the
physical volume shrinks to zero. 

In the nonnegative Yamabe classes, where $R_0 \ge 0$, one can just discard the $R_0$ term
in inequality (\ref{in'}) to get
\begin{equation}
 K^{ij}_{TT}K^{TT}_{ij}(max) \ge  {2 \over 3}K^2\phi^{12}_{max}.
\label{in''}
\end{equation}
From this one gets
\begin{equation}
V_p = \int\phi^6dv \le \phi^6_{max}V_0 =\sqrt{3 K^{ij}_{TT}K^{TT}_{ij}(max) \over
2K^2}V_0.
\end{equation}

This estimate holds always in the positive and zero Yamabe classes; in particular, it
shows how the volume goes to zero as $K^2$ becomes large. In the negative Yamabe class
such an estimate will continue to hold, but only in the large $K^2$ limit. This can be
seen by looking at the polynomial on the right hand side of (\ref{in'}) assuming $R_0 =
-C^2$. 

 The final question is what happens when $K^2$ shrinks to zero. While we know
that the conformal factor increases and drags the volume with it, we have no guarantee
that the volume becomes infinitely large. The Lichnerowicz - York equation, when $K^2$
equals zero, reduces to the original Lichnerowicz equation. Therefore we know that the
Yamabe constant of the base metric must be the determining factor. If the base metric is
in the positive Yamabe class, then the Lichnerowicz equation has a regular solution with
finite physical volume, and a maximal slice emerges. This will form the limiting solution
to the sequence of Lichnerowicz - York equations. The volume of the maximal solution is
the largest volume we can specify. It, of course, depends both on the choice of base
metric and on the specified TT tensor. 

If the Yamabe class of the base metric is nonpositive, then no regular solution to the
Lichnerowicz equation exists. The conformal factor blows up as $K^2$ goes towards zero,
and the physical volume becomes unboundedly large.

Let us consider the case where the Yamabe constant is negative. This means that one can
make a conformal transformation to a metric where the scalar curvature is constant and
negative, i.e., $R_0 = -C$ = constant. One could pick the constant to equal $-1$, but
this is not important. This base metric will have total volume $V_0$. As usual, one
solves the Lichnerowicz - York equation for the conformal factor $\phi$, and calculates
the physical volume $V_p$. I will show that
\begin{equation}
V_p > \left({3C \over 2K^2}\right)^{3 \over 2}V_0 = \left({3|Y| \over 2K^2}\right)^{3
\over 2}. \label{V-p} 
\end{equation}
The equality in Eq.(\ref{V-p}) follows from the definition of the Yamabe constant,
Eq.(\ref{Y'}). This guarantees that the volume diverges as
$K^2
\rightarrow 0$.

Start with a set of data, a metric with constant negative scalar curvature, a TT
tensor, and $K^2$. First make a conformal transformation with a constant conformal
factor $\phi_0 = (3C/2K^2)^{1/4}$. This transforms the base manifold to one where
the scalar curvature $\overline{R} = -2K^2/3$ and the volume $\overline{V} =
(3C/2K^2)^{3/2}V_0$. The Lichnerowicz - York equation in this frame becomes
\begin{equation}
8\overline{\nabla}^2\overline{\phi} + {2 \over 3} K^2 \overline{\phi} +
\overline{K^{TT}}\overline{K_{TT}}\bar{\phi}^{-7} - {2 \over 3} K^2
\overline{\phi}^5 = 0,
\end{equation}
where $\overline{K^{TT}}\overline{K_{TT}} = K^{TT}K_{TT}(2K^2/3C)^3$. From the min-max
principle we know that at the minimum of $\overline{\phi}$ we have
$\overline{\nabla}^2\overline{\phi}
\ge 0$. Therefore we have
\begin{equation}
 {2 \over 3} K^2 (\overline{\phi}_{min}^5 - \overline{\phi}_{min})
- \overline{K^{TT}}\overline{K_{TT}}\bar{\phi}^{-7}_{min} \ge 0.
\end{equation}
This guarantees that $\overline{\phi}_{min} \ge 1$ and thus $\overline{\phi} \ge 1$. In
turn, this means that the physical volume is greater than $\overline{V} =
(3C/2K^2)^{3/2}V_0$.

We can do somewhat better than that by looking at the maximum of $\overline{\phi}$.
In this case we have
\begin{equation}
 {2 \over 3} K^2 (\overline{\phi}_{max}^5 - \overline{\phi}_{max})
- \overline{K^{TT}}\overline{K_{TT}}\bar{\phi}^{-7}_{max} \le 0.
\end{equation}
This can be rearranged to give 
\begin{equation}
   \overline{\phi}^{7}_{max}(\overline{\phi}_{max}^5 - \overline{\phi}_{max})
\le {3 \over 2}\overline{K^{TT}}\overline{K_{TT}}/K^2 = K^{TT}K_{TT}(2K^2/3)^2 {1 \over
C^3}.
\label{max} \end{equation}

This tells us that $\overline{\phi}_{max}$ cannot be very large. As $K^2 \rightarrow 0$,
the right hand side of Eq.(\ref{max}) shrinks to zero and $\overline{\phi}_{max}$
is pushed down to 1. Thus the volume of the physical space converges to the
volume of $\overline{V} = (3C/2K^2)^{3/2}V_0 = (3Y/2K^2)^{3/2}$. Another way of saying
this is that in the limit, as $K$ becomes small, the conformal factor becomes large,
which renders the TT tensor unimportant. The physical geometry then more and more
approximates a large-volume 3-manifold with constant negative scalar curvature
$\overline{R} = -2K^2/3$. An alternativeway of expressing this is that $V_pK^3 \approx
(3Y/2)^{3/2}$ = constant.

In summary: If the base metric is in the positive Yamabe class, and if we pick any
physical volume which is less than or equal to the volume of the maximal solution, there
exists a unique choice of $K^2$, which, on solving the Lichnerowicz - York equation,
generates a solution to the constraints with the chosen volume. If the base metric has
nonpositive Yamabe constant, one can choose any volume between 0 and $\infty$; then there
exists a unique choice of $K^2$ which delivers this volume, on solving the Lichnerowicz
- York equation.

\section{Other choices}
In addition to trading off between the physical volume and the value of $K$, there are
several other specifications of data for the Lichnerowicz - York equation that work.
Given a base metric, a TT tensor, and a constant
$K$, one can find a complete family of TT tensors by simply multiplying the given tensor
by a constant. Then, for each choice of the constant, one can solve the Lichnerowicz -
York equation and find the physical volume. One can specify the volume and solve for the
constant multiplying the TT tensor.
\subsection{Scaling the TT tensor}
We start off with a modified Lichnerowicz - York equation
\begin{equation}
8\nabla^2\phi - R\phi + \alpha^2K^{ij}_{TT}K^{TT}_{ij}\phi^{-7} - {2 \over 3}K^2\phi^5 =
0,
\label{ly'}
\end{equation}
and we let $\alpha^2$ range from 0 to $\infty$. We now look for the
relationship between the physical volume and the value of $\alpha^2$. Let us again
perform a perturbation calculation, i.e., we start with a solution to the Hamiltonian
constraint
\begin{equation}
R - K_{ij}^{TT}K^{ij}_{TT}  + {2 \over 3}K^2 = 0, \label{ham3}
\end{equation}
and perturb Eq.(\ref{ly'}) around it, holding the metric, the TT tensor, and $K^2$
fixed. We only perturb $\alpha^2$ and consider the change of $\phi$ away from $\phi
\equiv 1$. We get
\begin{equation}
8\nabla^2\delta\phi - R\delta\phi -7 K^{ij}_{TT}K^{TT}_{ij}\delta\phi - {10 \over
3}K^2\delta\phi = -\delta\alpha^2 K^{ij}_{TT}K^{TT}_{ij}.
\label{ly3}
\end{equation}

By using Eq.(\ref{ham3}) to eliminate $R$, we can simplify this equation to get
\begin{equation}
8\nabla^2\delta\phi  -(8 K^{ij}_{TT}K^{TT}_{ij} + {8 \over
3}K^2) \delta\phi = -\delta\alpha^2 K^{ij}_{TT}K^{TT}_{ij}.
\label{ly4} 
\end{equation}
 We can again use the min-max principle on this equation and if $\delta\alpha^2 > 0$ we
have
$\delta\phi > 0$. In this case, as distinct from the `changing $K^2$' case, as we
increase the $K^{TT}$ term the volume monotonically increases. This guarantees
uniqueness.

In order to investigate whether there is any restriction on the allowed range of the
volume, we need to look at what happens at the two extremes, i.e., when $\alpha^2
\rightarrow
\infty$ and $\alpha^2 \rightarrow 0$. 
Let us start with the case where $\alpha^2 \rightarrow
\infty$. The Hamiltonian constraint, after solving the equation, is
\begin{equation}
\overline{R} -\alpha^2 \phi^{-12} K_{ij}^{TT}K^{ij}_{TT}  + {2 \over 3}K^2 = 0.
\label{ham4}
\end{equation}
If $\phi$ were to remain bounded on the support of $K^{TT}$ as $\alpha^2 \rightarrow
\infty$, the central term in Eq.(\ref{ham4}) blows up while the other two terms remain
finite. This cannot happen; therefore $\phi$ must blow up on the support of $K^{TT}$, and
the volume of the physical space becomes unboundedly large. This happens with any choice
of base metric.

The other extreme is when $\alpha^2 \rightarrow 0$. In this case the TT term drops out
of the Hamiltonian constraint, and we get
\begin{equation}
\overline{R}   + {2 \over 3}K^2 = 0. \label{ham5}
\end{equation}
This is a manifold with constant negative scalar curvature, and this is only achievable
if the Yamabe constant of the base metric is negative. 

Therefore if the base metric is is in the negative Yamabe class, the conformal factor
tends to a finite limit and generates a manifold of constant negative scalar
curvature. This manifold of constant negative scalar curvature is the manifold of least
volume in the sequence, and any physical volume we pick must be at least as large as
this. We can find an expression for this minimum volume easily in terms of the Yamabe
constant. On a manifold of constant curvature the Yamabe constant is achieved with a
constant function. Therefore we have
\begin{equation}
{\overline{R}V_p \over V_p^{1 \over 3}} = -{2K^2 \over 3}V_p^{2 \over 3} = Y.
\end{equation}
From this we get
\begin{equation}
V_p = ({3|Y| \over 2K^2})^{3 \over 2}.\label{|Y|}
\end{equation}
This is the minimum volume we can specify if the Yamabe constant is negative. 

It is clear, from Eq.(\ref{|Y|}), that this mimimum volume shrinks to zero as $|Y|
\rightarrow 0$. We can, however, do somewhat better. Let us assume that we are given a
base metric in the positive Yamabe class. Let us make a  conformal transformation so as
to map the base metric to one with constant positive scalar curvature. Let us further
arrange that this constant $R_0 = 2K^2/3$. This is not really necessary, but it makes
the algebra somewhat easier. The L - Y equation now becomes
\begin{equation}
8\nabla^2 \phi - {2 \over 3}K^2\phi + \alpha^2 K^{TT}K_{TT} \phi^{-7} - {2 \over 3}K^2
\phi^5 = 0.
\end{equation}
At the maximum of $\phi$ we have that $\nabla^2\phi \le 0$.
This gives
\begin{equation}
\alpha^2 K^{TT}K_{TT} \ge {2 \over 3} K^2(\phi^8_{max} + \phi^{12}_{max}), \label{maxa}
\end{equation}
where, of course, the left hand side of Eq.(\ref{max}) is evaluated at the point where
$\phi$ achieves its maximum. First, we see that $\phi$ achieves its maximum on the
support of $K^{TT}$. Further, we can get an upper bound for $\phi_{max}$ in terms of the
maximum of $K^{TT}K_{TT}$ from
\begin{equation}
{3 \alpha^2 \over 2 K^2}(K^{TT}K_{TT})_{max} \ge \phi^8_{max}.\label{maxb}
\end{equation}
This is not in any way sharp. A `better' estimate can be obtained by turning
inequality (\ref{maxa}) into a cubic equation. This, however, offers no real improvement,
because we are really interested in the case when $\alpha \rightarrow 0$; since then
$\phi_{max} \rightarrow 0$ and the $\phi^{12}$ term can be neglected wrt the $\phi^8$
term.

From (\ref{maxb}) we get the following estimate for the physical volume
\begin{equation}
V_p = \int \phi^6 dv \le \phi^6_{max} \int dv =\phi^6_{max}V_0 \le 
\left[{3 \alpha^2 \over 2 K^2}(K^{TT}K_{TT})_{max}\right]^{3 \over 4} V_0, \label{maxc}
\end{equation}
where $V_0$ is the volume of the base, constant curvature, metric. This shows that $V_p$
goes to zero as $\alpha$ goes to zero like $\alpha^{3/2}$. One could replace $V_0$ in
Eq.(\ref{maxc}) by the Yamabe constant to give
\begin{equation}
V_p \le \alpha^{3 \over 2} \left[{3 \over 2K^2}\right]^{9 \over 4}
\left[(K^{TT}K_{TT})_{max}\right]^{3 \over 4}Y^{3 \over 2}.
\end{equation}

\subsection{Scaling the base metric}
A third way of interpreting the Lichnerowicz - York equation is to choose a base metric
$g_b$, a TT tensor, a physical volume $V_p$, and the trace of the extrinsic curvature
$K$, but to assume the base metric is defined only up to a multiplicative constant
$g'_b = \beta^4g_b$. The given TT tensor will still be TT wrt the adjusted metric. The
value of $\beta$ is a freely choosen parameter. 

The behaviour of this system exactly mirrors the second system analysed, where the TT
tensor is scaled. This is because the following global scaling
\begin{equation}
(g_b, K^{TT}_{ij}, K) \rightarrow (\beta^4g_b, \beta^{-2}K^{TT}_{ij}, K)
\end{equation}
leaves the physical metric, and also the physical volume, unchanged. Therefore there is a
tradeoff between scaling on $g_b$ and a scaling of $K^{TT}$. 
We have the following chain
\begin{equation}
(g_b, K^{TT}_{ij}, K) \rightarrow (\beta^4g_b, K^{TT}_{ij}, K)
\rightarrow (g_b, \beta^{2}K^{TT}_{ij}, K).
\end{equation}
The first link is the nontrivial change we wish to analyse. The second link is an
identity, which leaves $V_p$ unchanged. Therefore scaling the metric by $\beta^4$ is
exactly equivalent to scaling the TT tensor by $\beta^2$. 

This means that as $\beta \rightarrow \infty$, the physical volume monotonically
increases and becomes unboundedly large. As $\beta \rightarrow 0$, the behaviour depends
on the Yamabe number of the base metric. If the Yamabe number is negative the volume
shrinks to a finite volume, represented by a manifold of constant negative scalar
curvature, while if the Yamabe number is nonnegative, the manifold shrinks to zero
volume. The estimates given above in subsection A continue to hold, except that $\alpha$
has to be replaced by $\beta^2$.

\begin{acknowledgements}
For many years I have been indebted to the relativity group of the Institute of
Theoretical Physics of the Jagellonian University.  Since the first of my many visits, in
1987, I have found the Institute and its members to be welcoming and friendly as
well as providing a stimulating place in which to work. Much of the credit
for maintaining this environment is surely due to its director and senior academic,
Professor Andrzej Staruszkiewicz. I would like to dedicate this article to
Professor Staruszkiewicz on the occasion of his 65$^{th}$ birthday.
\end{acknowledgements}

\end{document}